\documentstyle[aps,preprint,cite,psfig]{revtex}
\tighten

\author{Tanja I. Sachse}
\title{
Generalized Heitler--London Theory for H$_3$: 
A Comparison of the Surface Integral Method with Perturbation Theory
}
\date{\today}
\address{Max-Planck-Institut f\"ur Str\"omungsforschung, Bunsenstr.10,
D-37073 G\"ottingen, Germany} 

\author{Ulrich Kleinekath\"ofer}
\address{Institut f\"ur Physik, Technische Universit\"at, 
D-09107 Chemnitz, Germany}

\begin{document}

\maketitle
\begin{abstract}
The generalized Heitler--London (GHL) theory provides a straightforward 
way to express the potential energy surface of H$_3$ in terms of Coulomb 
and exchange energies which can be calculated either by perturbation 
theory or using the surface integral method (SIM). By applying the 
Rayleigh--Schr\"odinger perturbation theory, GHL theory for the quartet spin 
state of H$_3$ is shown to yield results equivalent to the symmetrized 
Rayleigh--Schr\"odinger version of symmetry adapted perturbation theory 
(SAPT). This equivalence allows a comparison with the corresponding results 
obtained by the surface integral method. The surface integral result 
calculated with a product of atomic wave functions is found to have certain 
advantages over the perturbation approach. 
\end{abstract}
\newpage

\section{Introduction}

The generalized Heitler--London (GHL) theory provides a useful framework to
calculate the potential energy surfaces for polyatomic systems
\cite{kt1798,book,h3paper,uk00}. Since the potential energy is expressed in
terms of Coulomb and exchange energies it is possible to systematically
separate out many--body effects in every single term contributing to the
potential energy.  In this paper some aspects of the three--body exchange
effects occurring in H$_3$ are examined in more detail.
    
Axilrod, Teller and Muto \cite{axilrod} were the first to suggest a formula
describing the leading long range three--body dispersion term for three
spherically symmetric atoms. Since then the non--additive effects have been 
intensively studied and  several review articles have been published 
\cite{elro94,meat91,meat84}.  In the GHL approach the potentials can be 
decomposed into   Coulomb and exchange {\it  energies}, whereas in symmetry 
adapted perturbation theory (SAPT) these interactions are expressed in terms 
of Coulomb and exchange {\it integrals} in the manner first introduced by 
Heitler and London. Recently, SAPT was formulated for the interactions of 
trimers \cite{mos10395} and has been applied to numerical calculations up to 
third order for the quartet spin state of H$_3$ \cite{korona96} and for the 
helium--trimer \cite{lot00} up to third order. 
Other three--body calculations for H$_3$ are based on Heitler--London type 
calculations \cite{whea8495} and perturbation calculations making use of 
Uns\"old approximations \cite{zhan94}. In the former the splitting into 
Coulomb and exchange part is as pointed out by the author himself not 
completely rigorous.

In a previous paper \cite{h3paper} analytical results were reported for the
doublet as well as for the quartet spin state for the H$_3$ system based on
the GHL theory. Two kinds of exchange energies appear: cyclic exchange
energies, where all three electrons are involved, and two--body exchange
energies in the presence of the respective third atom. The cyclic exchange
energy of three hydrogen and three helium atoms \cite{surfpaper} was
calculated using the surface integral method (SIM) which was previously
applied to two atoms \cite{kt1798,book,kt9491,kt9993,ukleine10797,uk00}.
In a forthcoming paper \cite{perturb_kt} it will be demonstrated that all
exchange energies occurring in the H$_3$--system can be calculated either by
the surface integral method or by using perturbation theory, and the
corresponding results for the implicit three--body effect on the two--body
exchange energies will be derived and compared.

For H$_2$ it was previously shown that SAPT and GHL are equivalent
\cite{cwi19592}.  The purpose of this paper is to compare the surface
integral method calculations of the three--body effects in the exchange
energies based on an atomic product wave function with the results of first
to third order of SAPT which are only available for the quartet spin state
of H$_3$ \cite{korona96}. In order to perform this comparison it is
necessary to first prove that the SAPT and GHL theory expressions for the
energy of the quartet state are equivalent.  The results reveal that with
the zeroth order wave function the surface integral result contains parts
of the second order SAPT result and is therefore more efficient.

In Sections \ref{ghl} and \ref{polapp} the basic ideas of the GHL theory and 
polarization approximation are described. In Section \ref{comp} the 
equivalence of the GHL and the symmetrized Rayleigh--Schr\"odinger (SRS) 
theories is demonstrated order by order. The latter is designated a weak 
symmetry forcing SAPT. Section \ref{sim} reviews the surface integral 
method (SIM). Thereafter in Section \ref{res} the advantages of SIM over the 
perturbation approach will be demonstrated by comparing the numerical results 
of perturbation theory and SIM.

\section{Generalized Heitler--London theory for H$_3$}\label{ghl}   

The application of generalized Heitler--London theory to H$_3$ was previously 
discussed in Ref. \cite{h3paper}. The generalized Heitler--London equation is 
given by  
\begin{equation} 
\hat{H} F \;=\; \sum_{g} \epsilon_g \hat{T}(g) F 
\label{ghl_eq}
\end{equation} 
where $F$ is the localized, i.e. non--symmetrized wave function, $\hat{T}(g)$ 
designates a permutation operator for the electron coordinates, and   
$\epsilon_g$ stands for the Coulomb ($g = I$) and exchange energies 
($g \neq I$).  Applying results from the theory of the symmetric group, 
the energy eigenvalues of the Hamiltonian can be derived. For the 
H$_3$--system, the result for the two doublet states is  
\begin{eqnarray}
^{1/2}E_{GHL} 
&=& 
\epsilon_I \;- \epsilon_{123} \; \pm{}  \sqrt{\frac{1}{2} \;
\left[ (\epsilon_{12} - \epsilon_{23})^2 \;+ (\epsilon_{23} - \epsilon_{13})^2 
\;+ (\epsilon_{13} - \epsilon_{12})^2 \right]}  
\end{eqnarray} 
and for the quartet state 
\begin{eqnarray}
\label{v4}
^{3/2}E_{GHL} &=& \epsilon_I \;-\epsilon_{12} \;-\epsilon_{23} 
\;-\epsilon_{13} \;+ 2 \epsilon_{123} \; .
\end{eqnarray}
The remainder of this paper will be concerned only with the quartet state.

\section{Polarization approximation and generalized Heitler--London (GHL) 
theory}\label{polapp}

%In a previous paper \cite{perturb_kt} it was shown how the Coulomb and 
%exchange energies of GHL theory can be expressed in terms of 
%Rayleigh--Schr\"odinger perturbation theory.  
The Born--Oppenheimer 
non--relativistic Hamiltonian of the three--body system is given by  
\begin{equation} \label{h} \hat{H} = \hat{H}^0 +\hat{V} \end{equation} 
using \begin{eqnarray} \hat{H}^0 &=& \hat{H}^0_A +\hat{H}^0_B
+\hat{H}^0_C \\ \hat{V} &=& \hat{V}_{AB} +\hat{V}_{BC} +\hat{V}_{AC}   
\end{eqnarray} where $\hat{H}^0_A, \hat{H}^0_B$ and $\hat{H}^0_C$ are
the Hamiltonians of  three free hydrogen atoms and $\hat{V}_{AB},
\hat{V}_{BC}$ and $\hat{V}_{AC}$  describe the interaction between atoms
$A$ and $B$, $B$ and $C$, as well as  $A$ and  $C$, respectively.  The
polarization approximation \cite{hirsch} is based on the equation 
\begin{equation}  \hat{H} F \;=\; E_p F  \end{equation}  where the
polarization wave function $F$ and the polarization energy $E_p$ can be
written as perturbation series \begin{eqnarray} F &=& \sum \phi_n~, \\ E_p
  &=& \sum \epsilon_n~.  \end{eqnarray} The zeroth order polarization wave
function $\phi_0$ is the eigenfunction of the free Hamiltonian $\hat{H}^0$
and thus is a product of three free hydrogen wave functions.  Starting from
the GHL equation with $F$ chosen as the polarization wave function, Eq.
(\ref{ghl_eq}) together with the Hamiltonian Eq. (\ref{h})  can be
  written as
\begin{eqnarray} (\hat{H}^0 + \hat{V}) | \sum_n \phi_n \rangle  &=& 
\sum_g \epsilon_g \; \hat{T}(g) | \sum_{n=0}^N \phi_n \rangle \; . 
\label{ghl_per_h3}  \end{eqnarray}
Forming scalar products with
$\hat{T}(g) \phi_0$ for each group element $g$
  \begin{equation}
(\hat{T}(g) \phi_0, \; (\hat{H}^0+\hat{V}) \; \sum_{n=0} \phi_n) \;=\;  
\sum_{g'} \epsilon_{g'} \; (\hat{T}(g) \; \phi_0, \; \sum_{n=0}  
\hat{T}(g') \phi_n)  \end{equation}
 a system of linear equations can be
derived for the Coulomb energy  $\epsilon_I$ as well as for   the
exchange energies $\epsilon_g$ ($g \neq I$) in terms of Coulomb integrals
$J$,  exchange integrals $K_g$, and overlap integrals $S_g$: 
\begin{eqnarray} \left.  \begin{array} {r@{\quad:\quad}l} E_0 \; +J \;
\approx \; \epsilon_I \; +\sum_{g'\neq g} \epsilon_{g'} \;  S_{g^{'-1}}
& g = I \\  E_0 S_g \; +K_g \; \approx \; \epsilon_g \; +\sum_{g'\neq g}
\epsilon_{g'} \;  S_{g^{'-1}\, g} & g \neq I \end{array} \right. \; .
\label{jk} \end{eqnarray}
  The following notation for the $n$th order
overlap, Coulomb and exchange  integrals was used:  \begin{eqnarray} S_g
&:=& \sum_{n=0}^M S^n_g \label{sg}\\ J &:=& \sum_{n=0}^M J^n \label{j}\\
 K_g &:=& \sum_{n=0}^M K^n_g = \sum_{n=1}^M K^n_g \; , \label{kg} 
\end{eqnarray} where  
\begin{eqnarray} S^n_g &:=& (\hat{T}(g) \phi_0, \;
\phi_n) \label{sn} \\  J^{n} &:=& (\phi_0, \; \hat{V} \; \phi_{n-1})
\label{jn} \\  J^0 &=& E_0 \\ K^{n}_g &:=& (\phi_0, \; \hat{V} \;
\hat{T}(g^{-1}) \; \phi_{n-1}) \; . \label{kn} \end{eqnarray}
 The
equalities $S^n_{g^{-1}} = S^n_g$ and $K^n_{g^{-1}} = K^n_g$ hold.  In
Ref. \cite{perturb_kt} it will be shown how the Coulomb and exchange
energies  can be expressed in terms of Coulomb, exchange and overlap
integrals and how  the order--by--order contributions to the Coulomb and
exchange energies can  be found. 

The convergence properties of the polarization theory
have been extensively discussed for the case of two hydrogen atoms
\cite{jezrev}. For low orders it was shown that the perturbation series
rapidly converges to the Coulomb energy
\cite{jezrev,chal1177,cwi19592,kt16289} though this is not the limit
for the infinite order expansion. It is assumed that the behavior of this
perturbation theory for a system of two atoms also roughly holds in the
case of three atoms \cite{mos10395,korona96}. Since here we are only
interested in low orders, especially the first, this expected behavior
justifies approximating the localized wave function via the polarization
approximation for three hydrogen atoms as well.

\section{Equivalence of the GHL and SRS theory for quartet H$_3$}
\label{comp}
 
In this section the order--by--order equivalence of the complete energy
expressions obtained by using either the GHL or the SRS theory will be 
demonstrated. Both the GHL and SRS theories start with the Hamiltonian 
Eq. (\ref{h}) and a zeroth order wave function which is a product of three 
free hydrogen atom wave functions. 
To demonstrate the equivalence of the first order expressions the first 
order SRS term will be expressed in terms of Coulomb and exchange energies. 
In Eq. (12) of Ref. \cite{korona96} this term is given by 
\begin{eqnarray}
^{3/2}E^1_{SRS} &=& 
N_0 \; \Biggl[ <\psi_0| \; \hat{V} \; (1 -\hat{T}(12) -\hat{T}(23) 
-\hat{T}(13) +\hat{T}(123) +\hat{T}(132)) \; |\psi_0> \Biggr] \; , 
\end{eqnarray} 
which can be expressed with Eqs. (\ref{sn}) to (\ref{kn}) as 
\begin{eqnarray} 
^{3/2}E^1_{SRS} &=&  
N_0 \; \Biggl[ J^1 \;- K^1_{12} \;- K^1_{23} \;- K^1_{13} \;+ K^1_{123} 
\;+ K^1_{132} \Biggr] \; , \label{srs1}
\end{eqnarray} 
where
\begin{equation}
N_0 = 1 -S^0_{12} -S^0_{23} -S^0_{13} +S^0_{123} +S^0_{132} \; .
\label{n0}
\end{equation}

With Eq. (\ref{jk}) it is possible to express the first order contributions 
as   
\begin{eqnarray}
J^1 
&=& 
\epsilon^1_I \;+\epsilon^1_{12} S^0_{12} \;+\epsilon^1_{23} S^0_{23} 
\;+\epsilon^1_{13} S^0_{13} \;+\epsilon^1_{123} S^0_{123} 
\;+\epsilon^1_{132} S^0_{123} \\
K^1_{12} 
&=& 
\epsilon^1_{12} \;+\epsilon^1_I S^0_{12} \;+\epsilon^1_{23} S^0_{123} 
\;+\epsilon^1_{13} S^0_{123} \;+\epsilon^1_{123} S^0_{23} 
\;+\epsilon^1_{132} S^0_{13} \\
K^1_{23} 
&=& 
\epsilon^1_{23} \;+\epsilon^1_I S^0_{23} \;+\epsilon^1_{12} S^0_{123} 
\;+\epsilon^1_{13} S^0_{123} \;+\epsilon^1_{123} S^0_{13} 
\;+\epsilon^1_{132} S^0_{12} \\
K^1_{13} 
&=& 
\epsilon^1_{13} \;+\epsilon^1_I S^0_{13} \;+\epsilon^1_{12} S^0_{123} 
\;+\epsilon^1_{23} S^0_{123} \;+\epsilon^1_{123} S^0_{12} 
\;+\epsilon^1_{132} S^0_{23} \\
K^1_{123} 
&=& 
\epsilon^1_{123} \;+\epsilon^1_I S^0_{123} \;+\epsilon^1_{12} S^0_{23} 
\;+\epsilon^1_{23} S^0_{13} \;+\epsilon^1_{13} S^0_{12} 
\;+\epsilon^1_{132} S^0_{123} \\
K^1_{132} 
&=& 
\epsilon^1_{132} \;+\epsilon^1_I S^0_{123} \;+\epsilon^1_{12} S^0_{13} 
\;+\epsilon^1_{23} S^0_{12} \;+\epsilon^1_{13} S^0_{23} 
\;+\epsilon^1_{123} S^0_{123} 
\end{eqnarray}
On inserting into Eq. (\ref{srs1}) many terms cancel and Eq. (\ref{srs1}) 
is equivalent to the first order contribution to Eq. (\ref{v4})  
\begin{eqnarray}
^{3/2}E_{SRS}^1 &=&
N_0 \; \Biggl[ J^1 \;- K^1_{12} \;- K^1_{23} \;- K^1_{13} \;+ K^1_{123} 
\;+ K^1_{132} \Biggr] \nonumber\\
&=& 
\epsilon^1_I \;-\epsilon^1_{12} \;-\epsilon^1_{23} \;-\epsilon^1_{13} 
\;+\epsilon^1_{123} \;+\epsilon^1_{132} \;=\; ^{3/2}E_{GHL}^1 \; .
\label{equiv1}
\end{eqnarray}
The rest of the proof will be done by complete induction. The claim of the 
induction is the equivalence of the GHL and SRS energy expressions up to 
$n$th order. From Eq. (12) of \cite{korona96} the general $n$th--order 
expression for the interaction energy in SRS theory is found to be 
\begin{eqnarray}
\vphantom{E}^{3/2}E_{SRS}^n 
&=& 
N_0 \; \Biggl[ <\psi_0| \; \hat{V} \; (1 -\hat{T}(12) -\hat{T}(23) 
-\hat{T}(13) +\hat{T}(123) +\hat{T}(132)) \; |\psi_{pol}^{(n-1)}> \Biggr. 
\nonumber\\
&-& \Biggl.
\sum_{k=1}^{n-1} \; \vphantom{E}^{3/2}E_{SRS}^k \; <\psi_0| \; 
(1 -\hat{T}(12) -\hat{T}(23) -\hat{T}(13) +\hat{T}(123) +\hat{T}(132)) 
\; |\psi_{pol}^{(n-k)}> \Biggr] \nonumber\\
&=& 
N_0 \; \Biggl[ J^n\; -K^n_{12}\; -K^n_{23}\; -K^n_{13}\; +K^n_{123}\; 
+K^n_{132} \Biggr. \nonumber\\
&-& \Biggl. 
\sum_{k=1}^{n-1} \; \vphantom{E}^{3/2}E_{SRS}^k \; (-S^{n-k}_{12}\; 
-S^{n-k}_{23}\; -S^{n-k}_{13}\; +S^{n-k}_{123}\; +S^{n-k}_{132}) \Biggr] 
\label{srsn}
\end{eqnarray}
where $N_0$ is given by Eq. (\ref{n0}). Thus it is necessary to prove that 
\begin{eqnarray}
^{3/2}E_{GHL}^n 
&=& 
\epsilon^n_I \;-\epsilon^n_{12} \;-\epsilon^n_{23} \;-\epsilon^n_{13} 
\;+\epsilon^n_{123} \;+\epsilon^n_{132} \\ 
&=& 
\vphantom{E}^{3/2}E_{SRS}^n \; .
\end{eqnarray} 

To perform a proof by induction it is necessary to show that also the 
$(n+1)$st order terms of both theories are equal. To do so, the $(n+1)$st 
order of GHL theory is expressed in terms of the quantities occurring in SRS 
theory. This can be achieved by inserting the solutions of the set of linear 
equations Eq.~(\ref{jk}) 
 into the complete GHL energy for the 
H$_3$--quartet state \cite{footnote}
\begin{eqnarray}
^{3/2}E_{GHL}  
&=& 
\epsilon_I \;-\epsilon_{12} \;-\epsilon_{23} \;-\epsilon_{13} 
\;+\epsilon_{123} \;+\epsilon_{132} \\
&\approx& 
\sum_{n=0}^M \; ^{3/2}E_{GHL}^n 
\;=\; 
\sum_{n=0}^M \; \Bigl[ \epsilon_I^n\; 
-\epsilon_{12}^n\; -\epsilon_{23}^n\; -\epsilon_{13}^n\; +\epsilon_{123}^n\; 
+\epsilon_{132}^n \Bigr] \nonumber\\
&=& 
E_0 \; +\Bigl[ J\; -K_{12}\; -K_{23}\; -K_{13}\; +K_{123}\; +K_{132} \Bigr] 
\nonumber\\ 
&\vphantom{=}& \qquad\quad 
\Bigl[ 1\; -S_{12}\; -S_{23}\; -S_{13}\; +S_{123}\; +S_{132} \Bigr]^{-1} \;   
\end{eqnarray} 
where $J$, $K_g$, and $S_g$ have been defined in Eqs. (\ref{sg}) to (\ref{kg}).
To find the expression for the $(n+1)$st order contribution to the energy 
of the quartet state, the left hand side is first multiplied by the 
denominator 
\begin{eqnarray}
&\vphantom{=}&
\Bigl( \sum_{n=0}^M \; ^{3/2}E_{GHL}^n \Bigr) \; \Bigl[ 1\; -\sum_{n=0}^M 
(S^n_{12}\; +S^n_{23}\; +S^n_{13}) \; +\sum_{n=0}^M (S^n_{123}\; +S^n_{132}) 
\Bigr] \nonumber\\
&=& 
E_0 \; \Bigl[ 1\; -\sum_{n=0}^M (S^n_{12}\; +S^n_{23}\; +S^n_{13}) \; 
+\sum_{n=0}^M (S^n_{123}\; +S^n_{132}) \Bigr] \nonumber\\ 
&+& 
\sum_{n=0}^M \; \left[ J^n\; -K^n_{12}\; -K^n_{23}\; -K^n_{13}\; +K^n_{123}\; 
+K^n_{132} \right] \; .
\end{eqnarray}
Collecting terms of $(n+1)$st order leads to 
\begin{eqnarray}
&\vphantom{=}&
^{3/2}E_{GHL}^{n+1} \; (1\; -S^0_{12}\; -S^0_{23}\; -S^0_{13}\; +S^0_{123}\; 
+S^0_{132}) \nonumber\\
&=& 
J^{n+1}\; -K^{n+1}_{12}\; -K^{n+1}_{23}\; -K^{n+1}_{13}\; +K^{n+1}_{123}\; 
+K^{n+1}_{132}\; \nonumber \\
&&+E_0 \; (-S^{n+1}_{12}\; -S^{n+1}_{23}\; -S^{n+1}_{13}\; 
+S^{n+1}_{123}\; +S^{n+1}_{132}) \nonumber\\ 
&&- 
\sum_{k=0}^n \; ^{3/2}E_{GHL}^k \; (-S^{n+1-k}_{12}\; -S^{n+1-k}_{23}\; 
-S^{n+1-k}_{13}\; +S^{n+1-k}_{123}\; +S^{n+1-k}_{132})  
\end{eqnarray}
with the result that  
\begin{eqnarray}
^{3/2}E_{GHL}^{n+1} 
&=& 
N_0 \; \Biggl[ J^{n+1}\; -K^{n+1}_{12}\; -K^{n+1}_{23}\; -K^{n+1}_{13}\; 
+K^{n+1}_{123}\; +K^{n+1}_{132} 
\Biggr. \nonumber\\
&-& \Biggl. 
\sum_{k=1}^{n} \; E^{GHL,k}_{3/2} \; (-S^{n+1-k}_{12}\; -S^{n+1-k}_{23}\; 
-S^{n+1-k}_{13}\; +S^{n+1-k}_{123}\; +S^{n+1-k}_{132}) \Biggr] \; . 
\label{eghln1}
\end{eqnarray}
Using the claim of the proof, which stated that for all orders up to the 
$n$th the GHL term is equal to the SRS--term, $E^{GHL,k}_{3/2}$ in the last 
line can be replaced by $^{3/2}E^{(n+1)}_{SRS}$ for all orders $1, \ldots, n$. 
Thus Eq. (\ref{eghln1}) can be transformed into 
\begin{eqnarray}
^{3/2}E_{GHL}^{n+1} 
&=& 
N_0 \; \Biggl[ J^{n+1}\; -K^{n+1}_{12}\; -K^{n+1}_{23}\; -K^{n+1}_{13}\; 
+K^{n+1}_{123}\; +K^{n+1}_{132} 
\Biggr. \nonumber\\
&-& \Biggl. 
\sum_{k=1}^{n} \; \vphantom{E}^{3/2}E_{SRS}^k \; (-S^{n+1-k}_{12}\; 
-S^{n+1-k}_{23}\; -S^{n+1-k}_{13}\; +S^{n+1-k}_{123}\; +S^{n+1-k}_{132}) 
\Biggr] \\
&=& 
\vphantom{E}^{3/2}E_{SRS}^{n+1} 
\end{eqnarray}
and the equality also holds for the $(n+1)$st order. Thus the contributions 
to the energy of the H$_3$--quartet state in the SRS and GHL theories are 
equal order by order. 
\par\bigskip

One advantage of the GHL theory is that it permits the calculation of the 
exchange energies by other methods, such as the surface integral method. 
In Ref. \cite{korona96}, the non--additive energy terms of the quartet spin 
state of H$_3$ have been calculated up to third order. The first order terms 
can be split into a polarization and an exchange part. Since the first order 
polarization energy is pairwise additive, the only non--additive term in 
first order is contained in the exchange term which in Eqs. (23) and (55) of 
Ref. \cite{mos10395} is given by  
\begin{eqnarray} 
&\vphantom{=}&
E^1_{exch}(3,3) \;=\;  
<\psi_0| \; \hat{V}_{AB} \left( \hat{T}(23) +\hat{T}(13) +\hat{T}(123) 
+\hat{T}(132) \; -S^0_{23} -S^0_{13} -S^0_{123} -S^0_{132} \right)|\psi_0> 
\nonumber\\
&+& 
<\psi_0| \; \hat{V}_{AB} \left( \hat{T}(12) +\hat{T}(13) +\hat{T}(123) 
+\hat{T}(132) \; -S^0_{12} -S^0_{13} -S^0_{123} -S^0_{132} \right)|\psi_0> 
\nonumber\\
&+& 
<\psi_0| \; \hat{V}_{AB} \left( \hat{T}(12) +\hat{T}(23) +\hat{T}(123) 
+\hat{T}(132) \; -S^0_{12} -S^0_{23} -S^0_{123} -S^0_{132} \right)|\psi_0> 
\; , 
\end{eqnarray}
which can be expressed in terms of exchange energies as  
\begin{eqnarray} 
&\vphantom{=}&
E^1_{exch}(3,3) = \epsilon^1_{123} (1 -S^0_{123}) \; - \Bigl[ \epsilon^1_{12} 
(1 +S^0_{12}) -\epsilon^{H_2,1}_{12} (1 +S^0_{12}) \Bigr] \nonumber\\
&-& 
\Bigl[ \epsilon^1_{23} (1 +S^0_{23}) -\epsilon^{H_2,1}_{23} (1 +S^0_{23}) 
\Bigr] \; -\Bigl[ \epsilon^1_{13} (1 +S^0_{13}) -\epsilon^{H_2,1}_{13} 
(1 +S^0_{13}) \Bigr] \; .
\label{exna_1} 
\end{eqnarray} 
This term is also obtained if the pure two--body contributions are subtracted 
from Eq. (\ref{equiv1}).

\section{Surface integral method (SIM) for the calculation of exchange 
energies}\label{sim}

As shown in Refs. \cite{surfpaper} and \cite{perturb_kt} all exchange energies 
occurring in the GHL--description of the H$_3$ system, i.e. the two--body as 
well as the cyclic exchange energies, can be calculated by the surface 
integral method (SIM). The exchange energy $\epsilon_{g_0}$ associated with 
the arbitrary group element $g_0 \neq I$ is given accordingly by   
\begin{eqnarray}
\label{surfall}
\varepsilon_{g_0} 
&=& 
\Biggl[ \int_V dv \; \Bigl[ F^2 - (\hat{T}(g_0) F)^2 \Bigr] \Biggr]^{-1} \; 
\Biggl[ \frac{1}{2} \int\limits_{\Sigma}\left\{F \vec{\nabla}^9
\left[ \hat{T}(g_0) F\right] -\left[ \hat{T}(g_0) F\right]\vec{\nabla}^9 F 
\right\} \cdot{}  d\vec{s} \Biggr] \nonumber\\ 
&\vphantom{=}& \qquad\qquad\qquad  
-\sum_{g\ne I, g_0} \varepsilon_g \int\limits_V dv \left[ F (\hat{T}(g_0 g) F) 
\; - \; (\hat{T}(g_0) F) (\hat{T}(g) F) \right] \Biggr] \; .
\end{eqnarray}
In order to compare numerical results for three--body exchange effects with 
the published SAPT results for H$_3$ \cite{korona96}, an expression for the 
non--additive exchange energy has to be obtained using SIM. The 
non--additive exchange energy basically contains the cyclic exchange energy 
and the implicit three--body effects on the two--body exchange energies. As 
already pointed out in Ref. \cite{surfpaper} it can be shown that for a 
choice of the partial volume $V$ such that $F$ is localized inside, all 
quantities occurring in the sum of Eq. (\ref{surfall}) go to zero with at 
least a factor of $e^{-R}$ faster than the surface integral itself if all 
internuclear distances are larger or equal to $R$. This holds for all 
exchange energies. In a different paper \cite{perturb_kt} it will be shown how 
to find the implicit three--body effect from the complete surface integral 
expression for the two--body exchange energies. For product wave functions 
as used here the pure two--body part is given by the first line of formula 
Eq. (\ref{surfall}), i.e. surface integral (SI) over denominator. The 
implicit three--body effect is contained in the second line of Eq. 
(\ref{surfall}), i.e. the products of partial overlap integrals with 
exchange energies. Following the same scheme used in the Appendix of Ref. 
\cite{surfpaper}, these terms can be shown to asymptotically go to zero as 
$e^{-5 R}$ which is faster by a factor of $e^{-3 R}$ than the surface 
integral (SI) itself. 

Using these results a GHL non--additive exchange energy for the quartet state 
of H$_3$ can be defined by simply subtracting the pure two--body contribution 
from the two--body exchange energies 
%and the Coulomb energy from 
in the GHL result for the quartet state Eq. (\ref{v4}) 
\begin{equation}
  (^{3/2}E_{GHL})_{exch} = 2 \epsilon_{123}\; -\Bigl[
  \epsilon_{12} -\epsilon^{H_2}_{12} \Bigr] \; -\Bigl[\epsilon_{23}
  -\epsilon^{H_2}_{23} \Bigr] -\Bigl[\epsilon_{13} -\epsilon^{H_2}_{13}
  \Bigr] 
\label{exna_ghl}
\end{equation}
which can be calculated either by SIM or perturbation theory. 
The first order contribution to this non--additive term  
\begin{equation}
  (^{3/2}E^1_{GHL})_{exch} = 2 \epsilon^1_{123}\; -\Bigl[
  \epsilon^1_{12} -\epsilon^{H_2,1}_{12} \Bigr] \; -\Bigl[\epsilon^1_{23}
  -\epsilon^{H_2,1}_{23} \Bigr] -\Bigl[\epsilon^1_{13} -\epsilon^{H_2,1}_{13}
  \Bigr] 
\label{exna_ghl_1}
\end{equation}
differs from the respective SRS--term Eq. (\ref{exna_1}) only by overlap 
integrals that are negligible compared to one.  

A comparison of the numerical results of the first order non--additive
exchange energy Eq. (\ref{exna_1}) of SRS theory and the GHL term 
[Eq. (\ref{exna_ghl_1})] calculated by SIM using the zeroth order product 
wave function $F = 1/\pi^{3/2} \exp (-r_{1A}-r_{2B} -r_{3C})$ is given in 
Tables \ref{exna_equitab} and \ref{exna_tritab} and will be discussed in the 
next Section.

In summary,  the complete three--body exchange effect in H$_3$, which 
consists of the cyclic exchange energy and the effect of the presence of 
the third atom on the two--body exchange energies, can asymptotically be 
approximated by the surface integral for the cyclic exchange energy.

\section{Results}\label{res} 

In Tables \ref{exna_equitab} and \ref{exna_tritab} 
as well as Figures 1 and 2 the numerical results 
for the first order non--additive exchange energy of SRS theory are compared 
with three different SIM--terms: (i) the non--additive exchange energy of GHL 
theory Eq. (\ref{exna_ghl}), (ii) the cyclic exchange energy (complete SIM 
expression Eq. (\ref{surfall}) with overlaps),
 (iii) the surface integral (SI) of the cyclic 
exchange energy only (without overlaps). All these quantities have been calculated using the 
zeroth order localized wave function $F = 1/\pi^{3/2} \exp (-r_{1A} -r_{2B} 
-r_{3C})$. Since the exchange energies calculated by SIM cannot be given a 
definite perturbative order (due to the fact that only part of the complete
space is used in the calculation) the quantity (i) is not expected to 
yield the same numerical results as the first order non--additive exchange 
energy of SRS theory.  But since the same zeroth order product wave function 
was used to calculate both terms it is expected that both quantities exhibit 
a similar overall behavior in the range of parameters studied.

In Table \ref{exna_equitab} results for equilateral triangular geometry of
the nuclei ranging between $R = 4$ and $R = 10$ atomic units are listed. 
Generally, all terms calculated by SIM have smaller absolute values than the 
first order perturbative ones. At $R = 4$ a.u., the absolute value of the 
complete SIM term Eq. (\ref{exna_ghl}) is 27 \% below the SRS result Eq. 
(\ref{exna_1}), the cyclic exchange energy is 38 \% smaller, and only the 
surface integral of the cyclic exchange energy is 25 \% greater in absolute 
value. At $R = 10$ a.u., however, all three quantities calculated by SIM are 
no longer distinguishable and are only 6 \% below the SRS result.
 
In Table \ref{exna_tritab} the results for isosceles triangles with equal
sides of length of 6 a.u. and with angles $\gamma_B$ varying between
30$^{\circ}$ and 180$^{\circ}$ are shown. All quantities except for the
surface integral without overlaps exhibit a change of sign in the region
around 120$^{\circ}$ and 150$^{\circ}$. At 30$^{\circ}$, (i) the absolute
value of the SIM term Eq. (\ref{exna_ghl}) is 31 \% smaller than the SRS
result, (ii) the cyclic exchange energy is 41 \% smaller, and again (iii)
the surface integral of the cyclic exchange energy only is 13 \% greater in
absolute value. At 180$^{\circ}$ on the other hand, only the value for the
surface integral has the wrong sign, while both the other terms have become
indistinguishable and are now 35 \% greater in absolute value than the SRS
term. The differences between the numerical results for the quantities
compared in Tables \ref{exna_equitab} and \ref{exna_tritab} are, as already
pointed out, not due to numerical problems but due to the fact that the
quantities are different by definition.

From the Tables it appears that for triangular geometries of the nuclei and 
internuclear distances $R \ge 4$ a.u. the first order non--additive exchange 
energy for the quartet state of H$_3$ can be quite well approximated by the 
surface integral of the cyclic exchange energy. This was stated in Ref. 
\cite{surfpaper} and has now been explained by the fact that all the SIM 
approximations (see section \ref{sim} and in Ref. \cite{surfpaper}) hold in 
this region. 

In Tables \ref{exna_equi_ord} and \ref{exna_tri_ord} as well as Figures 1
and 2 higher orders of SRS theory are also taken into account and compared
with the complete GHL non--additive exchange energy Eq. (\ref{exna_ghl}) in
order to show that SIM goes beyond the first order of SRS theory. For
equilateral triangular geometries of the nuclei and internuclear distances
larger than 6 a.u. the results of GHL theory lie between the first order
SRS term and the sum of the first and second order terms, approaching the
first order term for increasing distances. At 6 a.u. GHL is very close to
the first plus second order of SRS, and even at 4 a.u. GHL is only 17 \%
below the total sum up to third order of SRS theory.

For isosceles structures of the nuclei with equal internuclear distances of
6 a.u. the advantage of SIM over the first order SRS theory is even more
apparent. Starting at 60$^{\circ}$, the GHL result is closer to the first
plus second order than to the first order SRS term.  The change of sign
occurs for the first order between 120$^{\circ}$ and 150$^{\circ}$ whereas
for all other terms already between 90$^{\circ}$ and 120$^{\circ}$. The
differences of the GHL to the first plus second order SRS term range from
0.4\% at 60$^{\circ}$ to 33\% at 120$^{\circ}$ and 10\% at 180$^{\circ}$.
At 30$^{\circ}$ the GHL result is again only 16\% smaller than the SRS term
with the third order term included.

The advantage of SIM over the perturbative approach is that the surface 
integral SI is easily calculated numerically, and  including the partial 
overlap terms provides part of the second order SRS contributions.

\section{Conclusions}

This paper demonstrates how the perturbation series consisting of Coulomb, 
exchange and overlap integrals can be used to express the Coulomb and exchange 
energies occurring in GHL theory. Combining the perturbation series with the 
GHL theory yields an energy expression for the quartet spin state equivalent 
to that of symmetrized Rayleigh--Schr\"odinger perturbation theory given in 
\cite{korona96}.

It is possible to evaluate the exchange energies using the surface integral
method (SIM). The SIM has the advantage that it derives from a clear physical
picture for the exchange process in terms of the electrons continuously
trading places. For the cyclic exchange energies this method has already
been described in detail in Ref. \cite{surfpaper}, and for the implicit
three--body effect on the two--body exchange energies it will be shown in
Ref. \cite{perturb_kt}.

The long range behavior of the three--body terms entering the two--body
exchange energies and of the partial overlap integrals --- multiplied by 
two--body exchange energies in the expression for the cyclic exchange energy 
in Eq. (\ref{surfall}) --- indicate that for large internuclear separations 
the surface integral for the cyclic exchange energy is sufficient to describe 
the non--additive contribution to the exchange part of the quartet spin state. 
The numerical results in Tables \ref{exna_equitab} and \ref{exna_tritab} 
confirm this conclusion.

\section{Acknowledgements}

We thank K. T. Tang and J. P. Toennies for helpful discussions. U. K.
gratefully acknowledges financial support from the DFG.

%\begin{table}[tb]
%    \begin{tabular}{c|ccc|c}
%      \multicolumn{4}{c}{non--additive exchange energy} \\ order of wave
%      function& perturbation theory& &surface integral method& order of
%      wave function\\ \hline 0& $E^1_{exch}$ & --- & &\\ & & $\cdots$
%      &${\mathcal E}^{3/2}_{ex,NA}$& 0 \\ 1 & $E^2_{exch}$ & --- & &\\ 
%    \end{tabular}
%\caption{}\label{diag_tab}
%\end{table}

\begin{table}[tb]
    \begin{tabular}{c|cccc}
& \multicolumn{4}{c}{$E^1_{exch} [E_h]$} \\
$R [a_0]$& SRS Eq. (\ref{exna_1}) & GHL Eq. (\ref{exna_ghl}) & $2\epsilon_{123}$ 
(SIM)& 2 SI  \\ \hline
4 &$-3.83\cdot{}  10^{-3}$ &$-2.79\cdot{}  10^{-3}$ &$-2.39\cdot{}  10^{-3}$ 
&$-4.21\cdot{}  10^{-3}$\\
5 & ---                  &$-4.31\cdot{}  10^{-4}$ &$-4.16\cdot{}  10^{-4}$   
&$-5.26\cdot{}  10^{-4}$\\
6 &$-5.90\cdot{}  10^{-5}$ &$-5.19\cdot{}  10^{-5}$ &$-5.15\cdot{}  10^{-5}$   
&$-5.70\cdot{}  10^{-5}$\\
7 &$-5.88\cdot{}  10^{-6}$ &$-5.32\cdot{}  10^{-6}$ &$-5.31\cdot{}  10^{-6}$   
&$-5.55\cdot{}  10^{-6}$\\
8 &$-5.33\cdot{}  10^{-7}$ &$-4.89\cdot{}  10^{-7}$ &$-4.89\cdot{}  10^{-7}$   
&$-4.98\cdot{}  10^{-7}$\\
10&$-3.6\cdot{}  10^{-9}$  &$-3.4\cdot{}  10^{-9}$   &$-3.4\cdot{}  10^{-9}$    
&$-3.4 \cdot{}  10^{-9}$\\
    \end{tabular}
\caption{Comparison of the numerical results for the first order non--additive 
  exchange energy of SRS--theory (SRS$_1$ Eq. (\ref{exna_1})) with a similar but
  still different quantity derived from GHL theory Eq. (\ref{exna_ghl}),
  with the cyclic exchange calculated by SIM ($2 \epsilon_{123}$ (SIM))
including overlaps, and with 
  the surface integral SI of the cyclic exchange energy without overlaps 
(2 SI). The
  nuclei form equilateral triangles with sides of lengths $R$.}
\label{exna_equitab}
\end{table}

\begin{table}[tb]
    \begin{tabular}{c|cccc}
& \multicolumn{4}{c}{$E^1_{exch} [E_h], R_{AB} =R_{BC} =6$ a.u.} \\
$\gamma_B$ [degrees]&SRS Eq. (\ref{exna_1}) & GHL Eq. (\ref{exna_ghl}) &
$2\epsilon_{123}$ (SIM)&2 SI \\ \hline
30 &$-3.75\cdot{}  10^{-4}$ &$-2.60\cdot{}  10^{-4}$ &$-2.23\cdot{}  10^{-4}$  
&$-4.25\cdot{}  10^{-4}$ \\
60 &$-5.90\cdot{}  10^{-5}$ &$-5.19\cdot{}  10^{-5}$ &$-5.15\cdot{}  10^{-5}$  
&$-5.70\cdot{}  10^{-5}$ \\ 
90 &$-7.40\cdot{}  10^{-6}$ &$-6.05\cdot{}  10^{-6}$ &$-6.03\cdot{}  10^{-6}$  
&$-7.95\cdot{}  10^{-6}$ \\
120&$-3.42\cdot{}  10^{-7}$ &$2.61\cdot{}  10^{-7}$  &$2.60\cdot{}  10^{-7}$   
&$-1.62\cdot{}  10^{-6}$ \\
150&$8.84\cdot{}  10^{-7}$  &$1.31\cdot{}  10^{-6}$  &$1.30\cdot{}  10^{-6}$   
&$-5.83\cdot{}  10^{-7}$ \\
180&$1.10\cdot{}  10^{-6}$  &$1.48\cdot{}  10^{-6}$  &$1.48\cdot{}  10^{-6}$   
&$-4.10\cdot{}  10^{-7}$
    \end{tabular}
\caption{Comparison of the numerical results of SRS--theory with the same 
quantities as in Table \protect\ref{exna_equitab}. The nuclei form isosceles 
triangles with two sides of lengths $R_{AB} = R_{BC} = 6$ a.u., $\gamma_B$ 
is the angle included.} \label{exna_tritab}
\end{table}

\begin{table}[tb]
    \begin{tabular}{c|cccc}
& \multicolumn{4}{c}{$E_{exch} [E_h]$} \\
$R [a_0]$& SRS$_1$ Eq. (\ref{exna_1}) & SRS$_2$ & SRS$_3$ & GHL Eq. (\ref{exna_ghl}) 
\\ \hline
4 &$-3.83\cdot{}  10^{-3}$ &$-3.60\cdot{}  10^{-3}$ &$-3.34\cdot{}  10^{-3}$ 
&$-2.79\cdot{}  10^{-3}$\\
6 &$-5.90\cdot{}  10^{-5}$ &$-5.21\cdot{}  10^{-5}$ &$-5.03\cdot{}  10^{-5}$   
&$-5.19\cdot{}  10^{-5}$\\
7 &$-5.88\cdot{}  10^{-6}$ &$-4.77\cdot{}  10^{-6}$ &$-4.62\cdot{}  10^{-6}$   
&$-5.32\cdot{}  10^{-6}$\\
8 &$-5.33\cdot{}  10^{-7}$ &$-3.71\cdot{}  10^{-7}$ &$-3.57\cdot{}  10^{-7}$   
&$-4.89\cdot{}  10^{-7}$\\
10&$-3.6\cdot{}  10^{-9}$  &$-0.7\cdot{}  10^{-9}$   &$-0.7\cdot{}  10^{-9}$    
&$-3.4 \cdot{}  10^{-9}$\\
    \end{tabular}
\caption{Comparison of the numerical results for the  non--additive exchange 
  energy in GHL theory (GHL Eq. (\ref{exna_ghl})) with the first order
  non--additive exchange energy of SRS--theory (SRS$_1$ Eq. (\ref{exna_1})), 
  with the SRS non--additive exchange energy up to second order
(SRS$_2$)  \protect\cite{korona96} , and with up to third order SRS$_3$
  \protect\cite{korona96} . The nuclei form equilateral triangles with 
  sides of lengths $R$.} \label{exna_equi_ord}
\end{table}

\begin{table}[tb]
    \begin{tabular}{c|cccc}
& \multicolumn{4}{c}{$E_{exch} [E_h], R_{AB} =R_{BC} =6$ a.u.} \\
$\gamma_B$ [degrees] & SRS$_1$ Eq. (\ref{exna_1}) & SRS$_2$ & SRS$_3$ & GHL Eq. 
(\ref{exna_ghl}) \\ \hline
30 &$-3.75\cdot{}  10^{-4}$ &$-3.33\cdot{}  10^{-4}$ &$-3.08\cdot{}  10^{-4}$  
&$-2.60\cdot{}  10^{-4}$ \\
60 &$-5.90\cdot{}  10^{-5}$ &$-5.21\cdot{}  10^{-5}$ &$-5.03\cdot{}  10^{-5}$  
&$-5.19\cdot{}  10^{-5}$ \\ 
90 &$-7.40\cdot{}  10^{-6}$ &$-5.67\cdot{}  10^{-6}$ &$-4.98\cdot{}  10^{-6}$  
&$-6.05\cdot{}  10^{-6}$ \\
120&$-3.42\cdot{}  10^{-7}$ &$3.88\cdot{}  10^{-7}$  &$9.02\cdot{}  10^{-7}$   
&$2.61\cdot{}  10^{-7}$ \\
150&$8.84\cdot{}  10^{-7}$  &$1.43\cdot{}  10^{-6}$  &$1.88\cdot{}  10^{-6}$   
&$1.31\cdot{}  10^{-6}$ \\
180&$1.10\cdot{}  10^{-6}$  &$1.63\cdot{}  10^{-6}$  &$2.07\cdot{}  10^{-6}$   
&$1.48\cdot{}  10^{-6}$
    \end{tabular}
\caption{Comparison of the numerical results of GHL--theory with the same 
quantities as in Table \protect\ref{exna_equi_ord}. The nuclei form isosceles 
triangles with two sides of lengths $R_{AB} = R_{BC} = 6$ a.u., $\gamma_B$ 
is the angle included.} \label{exna_tri_ord}
\end{table}

\begin{figure}
\caption{Comparison of different orders of the non--additive exchange energy 
in SRS theory with the GHL result (filled triangles) calculated with SIM from 
Eq.\ \protect (\ref{exna_ghl}) for equilateral triangles. The first order SRS 
contribution is denoted by circles, and with all terms up to second order 
by open triangles. The stars show twice the surface integral of the cyclic 
exchange energy.} 
\label{equifig} 
\psfig{figure=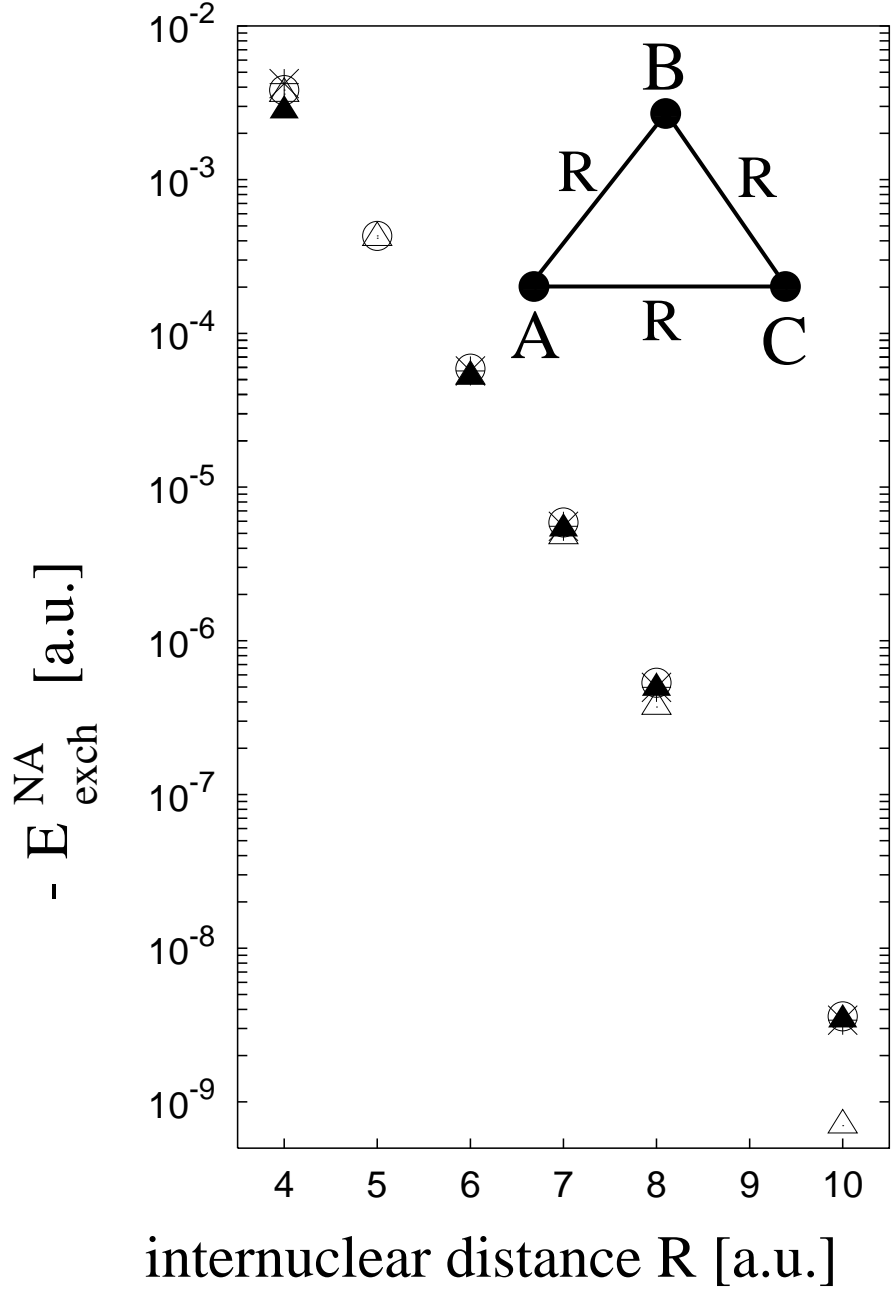,width=13cm}
\end{figure}

\begin{figure}
\caption{Comparison of different orders of the non--additive exchange energy 
in SRS theory with the GHL result (filled triangles) calculated with SIM from Eq.\ \protect 
(\ref{exna_ghl}) for isosceles triangles with $R_{AB} =R_{BC} =6$ a.u. 
as a function of the included angle $\gamma_B$. \protect\\ 
The first order SRS contribution is denoted by circles, and with all terms up 
to second order by open triangles. The stars show twice the surface integral of the 
cyclic exchange energy only. Note the change in the energy axis from linear
to logarithmic scale.}
\label{trifig} 
\psfig{figure=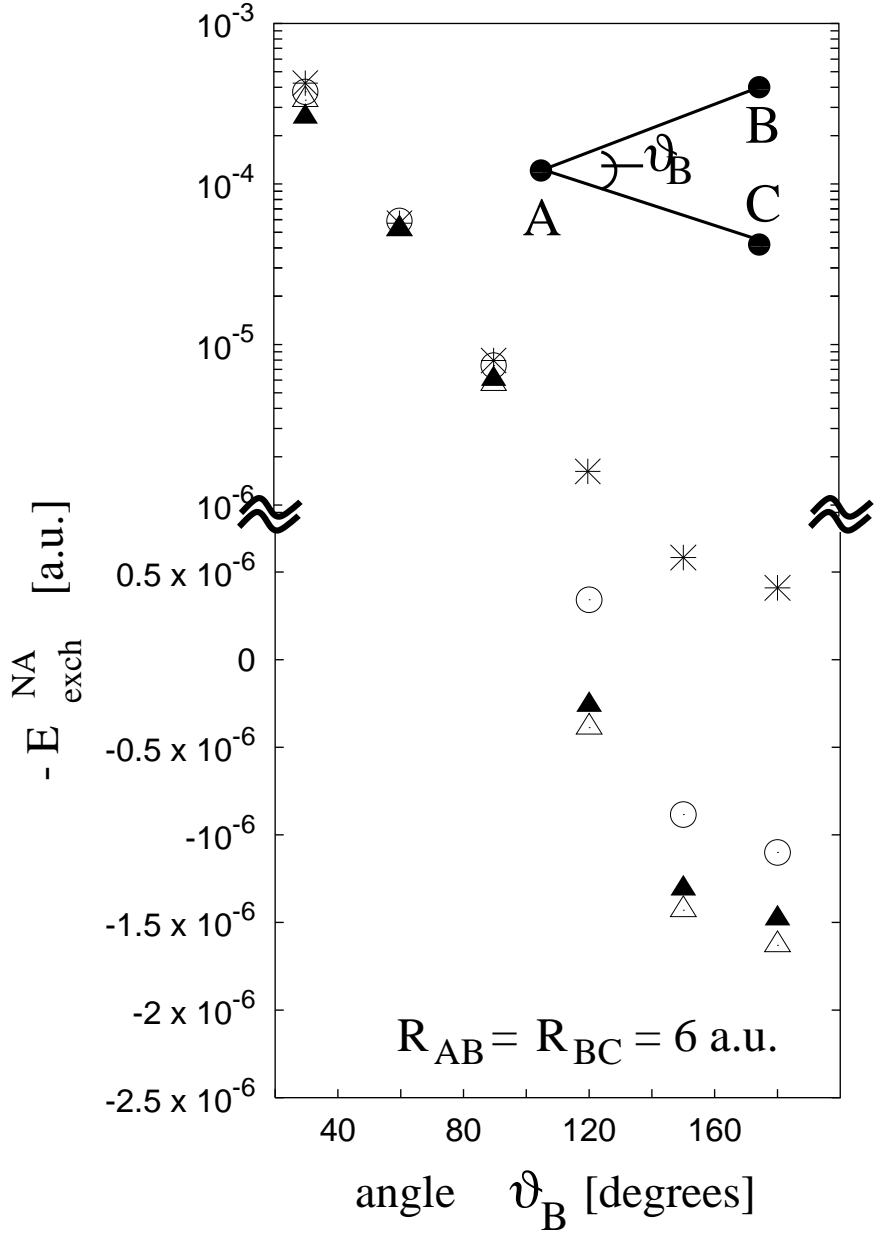,width=13cm}
\end{figure} 


\begin{thebibliography}{10}
\bibitem{footnote} The explicit expressions will be given in a 
forthcoming paper \protect \cite{perturb_kt}.
\end{thebibliography}
\end{document}